\documentclass[10pt]{article} 
\usepackage{amsmath}
\usepackage{amsfonts}
\usepackage{times}
\usepackage{epsf}
\def\({\left(}
\def\){\right)}
\def\[{\left[}
\def\]{\right]}
\def\non{ \nonumber }
\def\J{J(t)}
\def\X{\mathcal{X}}
\def\ga{\gamma}

\def\P{\mathcal{P}}
\def\T{\mathcal{T}}
\def\A{\mathcal{A}}
\def\M{\mathcal{M}}
\def\Ja{J_{\text{aff}}(t)}
\def\at{\mbox{\boldmath $\widetilde{a}$}}
\def\bt{\mbox{\boldmath $\widetilde{b}$}}
\def\ct{\mbox{\boldmath $\widetilde{c}$}}
\def\dt{\mbox{\boldmath $\widetilde{d}$}}
\def\mt{\mbox{\boldmath $\widetilde{m}$}}
\def\MT{\mbox{\boldmath $\widetilde{M}$}}
\def\qqq{\mbox{\boldmath $\mathcal{Q}$}}
\def\qq{\mathcal{Q}}
\def\ll{\mathcal{L}}
\def\lll{\widehat{\mathcal{L}}}
\def\aa{\mbox{\boldmath $a$}}
\def\bb{\mbox{\boldmath $b$}}
\def\cc{\mbox{\boldmath $c$}}
\def\dd{\mbox{\boldmath $d$}}
\def\ff{\mbox{\boldmath $f$}}
\def\hh{\mbox{\boldmath $h$}}
\def\tt{\mbox{\boldmath $t$}}
\def\zz{\mbox{\boldmath $z$}}
\def\ww{\mbox{\boldmath $w$}}
\def\mm{\mbox{\boldmath $m$}}

\def\V{\mathcal{V}}
                             
\def\BB{\mbox{\boldmath $B$}}
\def\ZZ{\mbox{\boldmath $Z$}}
\def\zzeta{\mbox{\boldmath $\zeta$}}
\def\TT{\mbox{\boldmath $T$}}

\begin{document} 

\rightline{LPTHE-00-04}
\vskip 1cm
\centerline{\LARGE Dual Baxter equations
and}
\centerline{\LARGE 
quantization of Affine Jacobian.} 
\vskip 2cm
\centerline{\large 
F.A. Smirnov 
\footnote[0]{Membre du CNRS}}
\vskip1cm
\centerline{ Laboratoire de Physique Th\'eorique et Hautes
Energies \footnote[1]{\it Laboratoire associ\'e au CNRS.}}
\centerline{ Universit\'e Pierre et Marie Curie, Tour 16, 1$^{er}$
		\'etage, 4 place Jussieu}
\centerline{75252 Paris Cedex 05, France}
\vskip2cm
\noindent
{\bf Abstract.} A quantum integrable model 
is considered which 
describes a quantization of affine hyper-elliptic
Jacobian. This model is shown to possess the
property of duality: a dual model with inverse
Planck constant exists such that the eigen-functions of its
Hamiltonians coincide with the eigen-functions of
Hamiltonians of the original model.
We explain that this duality can be considered as duality
between homologies and cohomologies
of quantized affine hyper-elliptic Jacobian.

\newpage

\section{Introduction.}

The relation between the affine Jacobian and integrable
models is well known ({\it cf.} \cite{mum}).
In the paper \cite{sn} we have shown that the
algebra of functions on the affine Jacobian is
generated by action of hamiltonian vector fields
from finite number of functions.
The latter functions are coefficients of
highest non-vanishing
cohomologies of the affine Jacobian. 
Actually, the idea that such description of the 
algebra of functions is possible appeared
from the paper \cite{toda} which
considers the structure of the algebra of
observables for quantum and classical Toda chain.

In the present paper we give quantum version of 
\cite{sn}. A quantum mechanical model
is formulated which gives a quantization of
the affine Jacobian. As usual in Quantum
Mechanics we can describe not the variety
itself but the algebra of functions on it (observables).
We need to show that the
quantum algebra of observables possesses
essential property of corresponding classical
algebra of functions. In our case 
this property is the possibility of creating
every observable from a finite number of
observables (cohomologies) by action
of Hamiltonians. 

In the process of realization of this program
we find the Baxter equations which
describe the spectrum of the model.
It happens that these equations possess 
the property of duality:
there is dual model with inverse Planck
constant for which the eigen-vectors are the same.
The algebras of observables of two
dual models commute.
The next ingredient of
our study is the method of separation of
variables developed by Sklyanin \cite{skl}.
Using this method we present the matrix elements
of any observable in terms of certain
integrals.

We show that the integrals in question are expresses
in terms of deformed Abelian integrals ({\it cf.}\cite{toda,abel}).
The observables for both dual
models are defined in terms of cohomologies.
The most beautiful feature of our construction is
that in these cohomologies enter the integrals for
matrix elements in such a way that the cohomologies
of dual model play role of homologies for original
one and vise a versa. We consider this relation
between week-strong duality in quantum theory
with duality between homologies and cohomologies as 
the most important conclusion of this paper.

\section{Affine Jacobian.}

In this section we briefly summarize necessary
facts concerning relation between integrable models and
algebraic geometry following the paper \cite{sn}.
The reason for repeating certain facts from
\cite{sn} is that we shall need them in slightly different
situation.

Consider $2\times 2$ matrix which depends polynomially on
the parameter $z$:
\begin{align}
&m(z)=\pmatrix a(z) &b(z)\non\\c(z) &d(z) \endpmatrix
\non
\end{align}
where the matrix elements are polynomials of the form:
\begin{align}
&a(z)=z^{g+1}+a_1z^{g}+\cdots +a_{g+1},\label{vii}\\
&b(z)=z^{g}+b_1z^{g-1}+\cdots +b_{g},\non\\
&c(z)=c_2z^{g}+c_3z^{g-1}\cdots +c_{g+2} ,\non\\
&d(z)=d_2z^{g-1}+d_3z^{g-2}\cdots +d_{g+1}
\non
\end{align}
In the the affine space $\mathbb{C}^{4g+2}$ with coordinates
$a_1,\cdots ,a_{g+1}$, $b_1,\cdots ,b_{g}$,
$c_2,\cdots ,c_{g+2}$, 
$d_2,\cdots ,d_{g+1}$
consider the (2g+1)-dimensional affine variety
$\M$
defined as quadric
\begin{align}
&f(z)\equiv a(z)d(z)-b(z)c(z)=1
\label{viii}
\end{align}
We consider this simplest situation, but
in principle
it is possible to put 
arbitrary polynomial
of degree $2g$ in RHS.

On the quadric $\M$ let us consider the sections
$\Ja $
defined by the equations: 
\begin{align}
&a(z)+d(z)=t(z)
\label{trace}
\end{align}
where 
$t(z)$ is given polynomial of the form:
\begin{align}
&t(z)=z^{g+1} +z^{g}t_1+\cdots +t_{g+1},\label{i}
\end{align}
The notation $\Ja $ stands for affine Jacobi variety.
The definition of  affine Jacobi variety and  its equivalence to $\Ja $
described above are given in the Appendix A. We include 
Appendix A because there is minor difference with the situation
considered in \cite{mum} and \cite{sn}.
The variety $\M$ 
is foliated into the affine Jacobians
$\Ja $. Mechanical
model described below provides a clever way of describing this
foliation.

We would like to understand the geometrical
meaning of quantum integrable models.
The general philosophy teaches that in order
to describe the quantization of
a manifold one has to deform the algebra
of functions on this manifold preserving
certain essential properties of this
algebra. The classical algebra must allow the
Poisson structure in order that quantization is possible.

Certain Poisson brackets for the coefficients
of matrix $m(z)$ can be introduced.
We do not write them down explicitly, if needed
they can be obtained taking classical limit of 
the commutation relations (\ref{comm}).
The algebra 
$$\widehat{\A}=\mathbb{C}\ [\ a_1,\cdots ,a_{g+1}, b_1,\cdots ,b_{g},
c_2,\cdots ,c_{g+2},
d_2,\cdots ,d_{g+1}]$$
becomes a Poisson algebra.
The most important properties of this Poisson structure are
the following.
First, 
the coefficients of the determinant $f(z)$ belong to the center
of the Poisson algebra, so, the equation
(\ref{viii}) is consistent with Poisson structure.
Second, the trace $t(z)$ generates commutative
sub-algebra:
$$\{t(z),t(z')\}=0$$
It can be shown, actually, that the coefficient $t_{g+1}$
of the trace belongs to the center, 
it is
convenient to put $t_{g+1}=(-1)^{g+1}2$. 
The subject of our study is the algebra of functions
on the $\M$:
$$\A=\frac{\widehat{\A}}
{\{f(z)=1,\ t_{g+1}=(-1)^{g+1}2\}}$$
on which the Poisson structure is well defined.

The
Poisson commutative algebra generated by the
coefficients
$t_1,\cdots ,t_ g$ is called the algebra 
of integrals of motion.
Introduce the commuting vector-fields
$$\partial _ i g=\{t_i,g\},\qquad i=1,\cdots g$$
The vector-fields $\partial _j$ describe motion along the sub-varieties
$\Ja $. 

One can think of these vector-fields as 
$$\partial _j=
\frac{\partial}{\partial \tau _j} $$
where $\tau _j$ are "times" corresponding
to the integrals of motion $t_j$.
Define the ring of integrals of motion
\begin{align}
&\T=\mathbb{C}\ [t _1,\cdots ,t_g]
\label{T}
\end{align}
Introduce the space of differential forms
$C ^k$ with
basis 
$$x\ d\tau _{i_1}\wedge\cdots\wedge d\tau _{i_k},\qquad
x\in \A$$
and the differential
$$d=\partial _jd\tau _j$$
Consider corresponding cohomologies $H^k$.
In the paper \cite{sn} the arguments are given
in favour of the following
\vskip 0.3cm
\noindent
{\bf Conjecture 1.}
{\it The cohomologies $H^k$ 
are finite-dimensional over the
ring $\T$,
they
are isomorphic to the cohomologies of the
affine variety} $\Ja $ {\it with $t$ in generic position.}
\vskip 0.3cm
\noindent
On the algebra $\A$ and on the spaces $C^k$
one can introduce degree \cite{sn}.
Take the basis of $H^g$ considered as
a vector space over $\T$ which is composed of homogeneous
representatives
$$\Omega _{\alpha}=g_{\alpha}
\ d\tau _{1}\wedge\cdots\wedge d\tau _{g}$$
where $\alpha $ takes finite number of values. 
The fact of foliation of $\M$ into varieties
$\Ja$ corresponds to the following statement
concerning the algebra $\A$ \cite{sn}
\vskip 0.3cm
\noindent
{\bf Proposition 1.}
{\it Every element $x$ of $\A$ can be presented
in the form}
\begin{align}
&x=\sum\limits _{\alpha}p_{\alpha}(\partial _1,\cdots ,\partial _g)
g_{\alpha}
\label{f=Ph}
\end{align}
{\it where $p_{\alpha}(\partial _1,\cdots ,\partial _g)$ are polynomials
of $\partial _1,\cdots ,\partial _g$ with coefficients in $\T$.}
\vskip 0.3cm
\noindent
The representation (\ref{f=Ph}) is not unique,
the equations 
\begin{align}
&\sum\limits _{\alpha} p_{\alpha}(\partial _1,\cdots ,\partial _g)
g_{\alpha}=0
\label{Ph=0}
\end{align}
are counted by $H^{g-1}$ \cite{sn}.

The formula (\ref{f=Ph}) can be useful only if we are able
to control the cohomologies. 
Concerning these cohomologies we adopt several
conjectures following \cite{sn}.

\section{Conjectured form of cohomologies.}

The affine variety $\Ja$ allows the following
description.
Consider the hyper-elliptic curve $X$ of genus $g$:
\begin{align}
&w^2 -t(z)w+1
=0
\label{curve}
\end{align}
This  curve has two points over the point $z=\infty $
which we denote by $\infty ^{\pm}$.

Consider a matrix $m(z)$ satisfying (\ref{viii}).
Take the zeros of $b(z)$:
$$b(z)=\prod\limits _{j=1}^g(z-z_j)$$
and
$$w_j=d(z_j)$$
Obviously $z_j,w_j$ satisfy the equation
of the curve $X$ (\ref{curve}).
Thus $m(z)$ defines a point $\P$ (divisor) 
on the symmetrized $g$-th power $X[g]$ of the curve $X$.
The divisor $\P$ consists of  
the points $p_j=(z_j,w_j)\in X$.

Oppositely one can reconstruct $m(z)$
starting form the divisor $\P$. Corresponding map is singular,
the singularities being located on
\begin{align}
&D=\{\P\ |\ p_i=\sigma (p_j)
\ \text{for some} \ i,j\ \text{or} \ p_i=\infty ^{\pm}
\ \text{for some} \ i\}
\label{div}
\end{align}
Where $\sigma $ is hyper-elliptic involution.
Thus the alternative description of $\Ja $ is 
$$\Ja =X[g]-D$$

Consider the meromorphic
differentials on $X$ with singularities at $\infty ^{\pm}$.
We chose the following basis of these differentials:
\begin{align}
&\mu _k(p)= {z ^{g+k}}\ \frac{dz}{y}, \quad
-g\le k\le 0\non\\
&\mu _k(p)=\bigl[y\frac {d}{dz}(z^{k-g-1}y)\bigr]_{\ge}
\ \frac{dz}{y},
\quad k\ge 1\label{abd}
\end{align}
where 
$p=(z,w)$, $\  y=2w-t(z)$, $[\ ]_{\ge}$ means that only
non-negative degrees of Laurent series in the
brackets are taken.

The form
$$\widetilde{\mu}_k=\sum\limits _i \mu _k(p_i)$$
is viewed as a form on $\Ja $.
It is easy to see that the forms $\mu _k$ (hence $\widetilde{\mu}_k$)
with $k\ge g+1$ are exact.
Consider the space $W^m$ with the basis:
$$\Omega _{k_1,\cdots ,k_m}=\widetilde{\mu}_{k_1}\wedge
\cdots\wedge \widetilde{\mu}_{k_m}$$
where $-g\le k_j\le g$.
Following \cite{sn} we adopt the
\vskip 0.3cm
\noindent
{\bf Conjecture 2.}
{\it We have}
\begin{align}
&H^m=\frac{W^m}{\sigma \wedge W^{m-2}}\label{H^m}
\end{align}
{\it where} 
$$\sigma =\sum\limits _{j=1}^g \widetilde{\mu}_j\wedge\widetilde{\mu}_{-j}$$
According to 
(\ref{div}) the singularities of differential 
forms occur either
at $p_i=\sigma (p_j)$ or at $p_i=\infty ^{\pm }$.
The non-trivial essence of  Conjecture 2 is that 
the first kind of singularities can be eliminated 
by adding exact forms. 
There are $(g-1)$-forms singular at $p_i=\sigma (p_j)$
such that these singularities disappear after
applying $d$.
This is the origin of the space
$\sigma \wedge W^{k-2}$ \cite{sn}.

Consider briefly the dual picture.
On the affine curve with punctures at $\infty ^{\pm}$
there are $2g+1$ non-trivial cycles $\delta _k$ with
$k=-g,\cdots ,g$. The cycles $\delta _k$ , $k<0$
are a-cycles, the cycles $\delta _k$ , $k>0$
are b-cycles and $\delta _0$ is the cycle around
$\infty ^+$. One defines the
cycles $\widetilde{\delta}_k$
on the symmetrical power of the affine curve.
The $\wedge$-operation is introduced for these cycles
by duality with cohomologies.
The non-trivial consequence of Conjecture 2
is that every cycle on $\Ja $ can be constructed
by wedging $\widetilde{\delta}_k$.
The formula dual to (\ref{H^m}) is
\begin{align}
&H_m=\frac{W _m}{\sigma ' \wedge W_{m-2}}\label{H_m}
\end{align}
where $W_m$ is spanned by
$$\Delta _{k_1,\cdots ,k_m}=\widetilde{\delta}_{k_1}\wedge
\cdots\wedge \widetilde{\delta}_{k_m}$$
and
$$\sigma '=\sum\limits _{j=1}^g\ \widetilde{\delta}_j\wedge
\widetilde{\delta}_{-j}
$$
We need to factorize over $\sigma ' \wedge W_{m-2}$
because the 2-cycle $\sigma '$ intersects with $D$.


Let us return to the relation of $H^g$ to the
algebra $\A$.
Notice that 
$$d\tau _1\wedge\cdots\wedge d\tau _g\simeq\widetilde{\mu}_1
\wedge\cdots\wedge\widetilde{\mu}_g\equiv \Omega
$$
The functions 
$$x_{k_1,\cdots ,k_g}=\Omega ^{-1}\ \Omega _{k_1,\cdots ,k_g}$$
are symmetric polynomials of
$z_1,\cdots ,z_g$. Recall that
$b_1,\cdots ,b_g$ are nothing but 
elementary symmetric polynomials of
$z_1,\cdots ,z_g$. 
Hence the coefficients of cohomologies have the form:
$$g_{\alpha}=g_{\alpha}(b_1,\cdots ,b_g)$$
The dimension of $H^g$ is determined by Conjecture 2:
$$\alpha =1,\cdots ,
\textstyle{\binom{2g+1}{g}-\binom{2g+1}{g-2}},$$
The equations (\ref{Ph=0}) are consequences of the following
ones
\begin{align}
\sum\limits_{k=1}^g \partial _k
\bigl(
\ \Omega ^{-1}
\ (\mu _{-k}\wedge \Omega _{k_1,\cdots ,k_{g-1}})\bigr)=0,
\quad \forall\ \Omega _{k_1,\cdots ,k_{g-1}}\in W^{g-1} 
\label{nuli}
\end{align}



\section{Quantization of affine Jacobian.}

Let us consider a quanization of algebra $\A$.
The parameter of deformation (Planck constant) is
denoted by $\gamma$, we shall also use
$$q=e^{i\gamma }$$
Consider the 2$\times$ 2 matrix $\mm (z)$ with
noncommuting entries. 
Suppose that the dependence on the
spectral parameter $z$ is
exactly the same as in classical case (\ref{vii}).
The variables $\aa _j$, $\bb _j$, $\cc _j$, $\dd _j$ 
are subject to commutation
relations which are summarized as follows:
\begin{align}
&r_{21}(z_1,z_2)\ \mm _1(z_1)\ k_{12}(z_1)
\ s_{12}\ \mm _2(z_2)\ k_{21}(z_2)=\non\\
&=\mm _2(z_2)\ k_{21}(z_2)
\ s_{21}\ \mm _1(z_1)\ k_{12}(z_1)\ r_{12}(z_1,z_2)
\label{comm}
\end{align}
where usual conventions are used: the equation (\ref{comm})
is written in the tensor 
product
$\mathbb{C}^2\otimes\mathbb{C}^2$,
$a_1=a\otimes I$, $a_2=I\otimes 2$, $a_{21}=Pa_{12}P$ where
$P$ is the operation of permuattions.
The $\mathbb{C}$-number matrices $r$, $k$, $s$ are:
\begin{align}
&r_{12}(z_1,z_2)={z_1-qz_2\over 1-q}\ (I\otimes I)+
{z_1+qz_2\over 1+q}\ (\sigma ^3\otimes \sigma ^3)
+\non\\&+2\ (z_1\sigma ^-\otimes \sigma ^++z_2\sigma ^+\otimes \sigma ^-),
\non\\
&k_{12}(z)=I\otimes(I-\sigma ^3)+
\(q^{-\sigma ^3}
+z(q^2-1)\sigma ^-
\)\otimes(I+\sigma ^3),\non\\
&
s_{12}=I\otimes I-(q-q^{-1})
\sigma ^-\otimes\sigma ^+
\label{rks}
\end{align}
These commutation relations
are important because they respect the form
of matrix $\mm (z)$ prescribed by (\ref{vii}), we
shall explain how they are related to more usual r-matrix relations
in the next section.

Define the polynomials:
\begin{align}
&\tt (z)=q\aa (z)+q^2 \dd (z)-z(q^2-1)\bb (z)\non\\
&\ff (z)=q\dd (z)\tt (zq^{-2})-q^2\dd (z)\dd (zq^{-2})-q\bb (z)\cc (zq^{-2})
\label{tf}
\end{align}
The algebra $\widehat{\A}(q)$ is generated by 
$\quad
\aa _1,\cdots ,\aa _{g+1}$, $\quad\bb _1,\cdots ,\bb _{g}$,
$\quad\cc _2,\cdots ,\cc _{g+2}$,
$\dd _2,\cdots ,\dd _{g+1}$,
The polynomial $\ff (z)$ belongs to the center of $\widehat{\A}(q)$.
The coefficients of $\tt (z)$ 
are commuting, actually, $\tt _{g+1}$
belongs to the center of $\widehat{\A}(q)$.
We define:
\begin{align}
\A (q)=\frac {\widehat{\A}(q)}{\{\ff (z)=1,\ \tt _{g+1}=(-1)^{g+1}2\}}
\non
\end{align}
The non-commutative algebra $\A(q)$ defines a
quantization of the algebra of function on the 
quadric $\M$. However, we cannot define directly
the quantization of the algebra of functions on
the affine Jacobian because the coefficients
of $\tt (z)$
are not in the center of $\A (q)$.
What we can do is to describe the quantum version of
Proposition 1 and of description of cohomologies. 
The exposition will be more detailed than in  the classical case.

Like in  the paper \cite{toda} we accept the following
\vskip 0.3cm
\noindent
{\bf Conjecture 3.}
{\it The algebra }$\A (q)$ {\it is spanned as linear space
by elements of the form:}
\begin{align}
x=p_L(\tt _1,\cdots ,\tt _g)
g(\bb _1,\cdots ,\bb _g)
p_R(\tt _1,\cdots ,\tt _g)
\label{f=PHP}
\end{align}
{\it where 
$p_L(\tt _1,\cdots ,\tt _g),\quad $ $g(\bb _1,\cdots ,\bb _g),\quad$
$p_R(\tt _1,\cdots ,\tt _g)$ are polynomials.}
\vskip 0.3cm
\noindent
We were not able to prove this statement, however,
since the algebra $\A (q)$ is graded we can check it
degree by degree. This has been done up to
degree 8. Notice the similarity between the representation 
(\ref{f=PHP})
and the representation for spin 
operators proved in \cite{maillet}. 
Conjecture 3 implies that certain generalization
of  the results of \cite{maillet} is possible.
In fact the formula (\ref{f=PHP}) is similar to
the formula (\ref{f=Ph}): we can either symmetrize or anti-symmetrize
$\tt _j$ in (\ref{f=PHP}) which corresponds in 
classics to multipilation by $t_j$
or to applying $\partial _j$.
In order 
to have complete agreement with clasical case we have to
show that only finitely many different polynomials $g(\bb _1,\cdots \bb _g)$
(cohomologies) 
create entier algebra $\A (q)$.

Notice that the commutation
relations (\ref{comm}) imply in particular that
$$[\bb (z), \bb (z ')]=0$$
which means that we have the commutative family of
operators $\zz _j$ defined by
$$\bb (z)=\prod\limits (z-\zz _j)$$
So, every polynomials $g(\bb _1, \cdots ,\bb _g)$
can be considered as symmetric polynomial
of $\zz _j$ and vice versa.

It is very convenient to use the following formal definitions.
Consider the ring $\T$ defined in (\ref{T}).
By $\V ^k$ we denote the space of anti-symmetric polynomials
of $k$ variables such that their degrees with respect to
every variable is not less than $1$ with coefficients in $\T\otimes\T$.
In other words $\V ^k$ is the space spanned by the polynomials:
$$
p_L\cdot h\cdot p_R\equiv p_L(t_1,\cdots , t_g)h(z_1,\cdots ,z_k)p_R(t_1',\cdots , t_g')
$$
where $h$ is anti-symmetric, vanishing when one of $z_j$ vanishes.
The following operations can
be defined.
\newline
1. Multiplication by $t_j$ and $t_j'$. 
\newline
2. Operation $\wedge \ :\ \V ^ k\otimes\V ^l\to\V ^{k+l}$ which
is defined 
as follows:
$$(p_L\cdot h\cdot p_R)\wedge
(p_L'\cdot h'\cdot p_R')=p_Lp_L'\cdot (h\wedge h')\cdot p_Rp_R'$$
where
\begin{align}
(h\wedge h')&(z_1,\cdots ,z_{k+l})=\non\\
&=\frac 1{k!\ l!}
\sum\limits _{\pi\in S_{k+l}}(-1)^{\pi}
\ h(z_{\pi(1)},\cdots ,z_{\pi(k)})
h'(z_{\pi(k+1)},\cdots ,z_{\pi(k+l)})
\non
\end{align}
 
We have a map
$$\V ^g\stackrel{\chi}{\longrightarrow } \A (q)$$
defined on the basis elements as
$$\chi (p_L\cdot h\cdot p_R)=
p_L(\tt _1,\cdots ,\tt _g)\ \frac {h(\zz _1,\cdots ,\zz _g)}
{\prod\zz _i\prod _{i<j}(\zz _i -\zz _j)}
\ p_R(\tt _1,\cdots ,\tt _g)$$ 
and continued linearly.
The Conjecture 3 states that this map is surjective.
We want to describe the kernel of the map $\chi $.

First, consider the space $\V ^1$. The elements of
this space are polynomials of
one variable $z$ with coefficients in $\T\otimes \T$.
In Appendix B we describe certain 
basis
in $\V ^1$ 
considered as a linear space over $\T\otimes \T$.
The basis in question consists of the 
polynomials: $s_{k}$ with $k\ge -g$
such that the degree of $s_{k}$ with respect to
$z$ equals $g+k+1$. The kernel of $\chi $
is the joint of three sub-spaces, let us describe them.
\vskip 0.2cm
\noindent 
1. For $k\ge g+1$
we have:
\begin{align}
\chi \bigl(s_k \wedge \V ^{g-1}\bigr)=0
\label{p1}
\end{align}
\vskip 0.2cm
\noindent 
2. Consider $c\in \V ^2$ defined as
$$c=\sum\limits _{j=1}^g\ s_j\wedge s _{-j}$$
we have
\begin{align}
\chi\bigl( c\wedge \V ^{g-2}\bigr)=0
\label{p2}
\end{align}
\vskip 0.2cm
\noindent 
3. Consider $d\in \V ^1$ defined as
$$d=(
t_j-t'_j)s_{-j}$$
we have
\begin{align}
\chi \bigl(d\wedge\V ^{k-1}\bigr)=0
\label{p3}
\end{align}

The construction of  the space 
$$\frac {\V ^g}{\text{Ker}(\chi)}\simeq \A (q)$$
is in complete correspondence with the classiccal case.
In classics we start with all the 1-forms $\widetilde{\mu} _k$. Imposing
(\ref{p1})  
corresponds to throwing away the exact forms
and working with $\widetilde{\mu} _k$ for  $k=-g\cdots , g$ only.
Imposing
(\ref{p2}) corresponds to
factorizing over $\sigma \wedge W^{k-2}$ in classics.
Finally, (\ref{p3})
corresponds to
the equation (\ref{nuli}).

The origin of the equations (\ref{p1}), (\ref{p2}), (\ref{p3})
will be explained in the Section 9. There should be
purely algebraic method of prooving these equations, but
we do not know it. It is important to mention that
accepting Conjecture 3 we are forced to 
conclude that the kernel of $\chi $ is completely
described by the equations (\ref{p1}), (\ref{p2}), (\ref{p3}).
This is proved by calculation of characters similarly to
that of \cite{toda}.

\section{ The realization of $\mathbf{A(q)}$.}

We want to describe a realization of the algebra
$\A (q)$ in a space of functions. 
Consider the quantum mechanical system
described by the operators $x_j$ with $j=1,\cdots , 2g+2$
and $y$ (zero mode).
The operators $x_j$ and $y$ are self-adjoint, they satisfy
the commutation relations:
\begin{align}
&x_kx_l=q^2x_lx_k\qquad k<l,\non\\
&yx_k=q ^2x_ky\qquad \forall k\non
\end{align}
The hamiltonian of the system is 
$$\hh =q^{-1}\sum _{k=1}^{2g+2}x_kx_{k-1}^{-1}$$
where 
$$ x_{2g+3}\equiv qyx_1$$
Physically this model defines the simplest lattice regularization
of the chiral Bose field with modified energy-momentum
tensor.

It is useful to double the
number of degrees of freedom.
Consider the algebra $A$ generated by
two operators $u$ and $v$ satisfying the commutation
relations:
$$uv =q vu$$
Take the algebra $A^{\otimes (2g+2)}$
the operators
$u_j$, $v_j$ ($j=1,\cdots ,2g+2$) are defined as $u$ and $v$
acting in $j$-th tensor component. 
The original operators $x_i$
are expressed in terms of $u_i,v_i$ as follows:
$$x_k=v_k\prod _{j=1}^{k-1}u_j^{-2},\quad y=\prod _{j=1}^{2g+2}u_j$$
Consider the ``monodromy matrix''
\begin{align}
&\mt (z)= \begin{pmatrix}
\at (z)&\bt (z)\\
\ct (z)&\dt (z)\end{pmatrix}
=l_{2g+2}(z)\ \cdots \ l_{1}(z)\label{mt}
\end{align}
where the l-operators are
\begin{align}
l(z)={1\over\sqrt{z}}
\begin{pmatrix}
z
u &
-qvu\\
zv^{-1}u^{-1} &0\end{pmatrix}
\end{align}
This is a particular case of more genaral l-operator
$l(z,\kappa)$ in which the last matrix
element is not $0$ but $\kappa z u $, the 
model corresponding to the
latter
l-operator is a  subject of study in a series of papers \cite{fv}.

The matrix elements of the matrix $\mm (z)$
satisfy the commutation relations:
\begin{align}
&r _{12} (z_1, z_2) \mt _1(z_{1})
 \mt  _2(z_{2})=
\mt  _2(z_{2})\mt _1(z_{1})r_{12}(z_1, z_2)
\label{rll}
\end{align}
where the r-matrix $r_{12}(z_1, z_2)$ is defined
ealier (\ref{rks}).
These are canonical r-matrix commutation relations.
The quantum determinant of the matrix 
$\mt (z)$ is defined by
$$\ff (z)=\dt(z)\at(zq^{-2})-\bt(z)\ct(zq^{-2}),$$
it belongs to the center, in our realization of
$\mt (z)$ one has $\ff (z)=1$.
The trace of  $\mt (z)$ generates
commuting quantities, we denote 
this trace as follows:
$$\at(z)+\dt(z)=y \tt (z)$$

The matrix elements of the matrix $\widetilde{m}(z)$
are of the form:
\begin{align}
&\at(z)=\at_0z^{g+1}+\at_1z^{g}+
\cdots +\at_{g+1},\label{abcd}\\
&\bt(z)=\bt _0z^{g}+\bt _1z^{g-1}+
\cdots +\bt_{g},\non\\
&\ct(z)=\ct_1z^{g+1}+\ct_2z^{g}+
\cdots +\ct_{g+1}z ,\non\\
&\dt(z)=\dt_1z^{g}+\dt_2z^{g-1}+
\cdots +\dt_{g}z
\non
\end{align}
where, in particular, $\at_0=y$. 
This form of polynomials $\at(z)$, $\bt(z)$, $\ct(z)$, $\dt(z)$
does not corresponds to what we have in the classical
model of affine Jacobian. This is the reason for modifying
the matrix $\widetilde{m}(z)$ as follows:
$$
\mm (z)=\begin{pmatrix}
\at_0\bt _0^{-1} &0\\
-\dt _1\bt _0^{-1}&1
\end{pmatrix}
\mt (z)
\begin{pmatrix}
\bt _0\at_0^{-1} &0\\
q\dt _1\at_0^{-1}&1
\end{pmatrix}
$$
The matrix elements of this matrix have structure (\ref{vii}),
they satisfy closed commutation relations (\ref{comm}),
the operators $\ff (z)$ and $\tt (z)$ defined for these two matrices 
coincide,
in particular we have
$$\tt _1=\hh$$
Thus the  modification of 
matrix $\mt (z)$ which is necessary
for relation to the affine Jacobian
is responcible for appearing of
strangely looking commutation relations (\ref{comm}).


\section
{Q-operator.}

Our first goal is to define Baxter's Q-operator.
Let us realize  the operators $v,u $ in $L_2(\mathbb{R})$
as follows
$$v=e^{\varphi },\qquad u=e^{i\ga{d\over d\varphi }}$$
We shall work in the $\varphi$-representation,
i.e. in the space 
$\mathfrak{H}=(L_2(\mathbb{R}))^{\otimes (2g+2)}$.
Following the standart procedure ({\it cf.} \cite{gp})
one introduces the vectors 
$Q(\zeta\ |\ \psi _1,\cdots ,\psi _{2g+2})$
which depend on 
$$\zeta =\textstyle{1\over 2}\log z$$ 
and $2g+2$ additional parameters, $\psi _j$,
and satisfy the equation:
\begin{align}
(-1)^{g+1}\tt (z)Q(\zeta\ &|\ \psi _1,\cdots ,\psi _{2g+2})=\non\\&=
Q(\zeta+i\ga \ |\ \psi _1,\cdots ,\psi _{2g+2})
+Q(\zeta-i\ga \ |\ \psi _1,\cdots ,\psi _{2g+2})
\non
\end{align}
In  $\varphi$-representation the ``components'' of these vectors are given
by
\begin{align}
Q(\varphi _{1},&\cdots ,\varphi _{2g+2}\ |
\ \zeta \ |\ \psi _1,\cdots ,\psi _{2g+2})=
\non\\&=
e^{{1\over 2}(1+{\pi\over\ga})\zeta+{1\over 4i\ga}\zeta ^2 }\prod _{k=1}^{2g+2}
\lambda (\zeta\ |\ \varphi _k-\psi _k\ )
\langle\ \varphi _k \ |\ \psi _{k-1}\ \rangle
\label{ker}
\end{align}
where $\psi _0\equiv \psi _{2g+2}$,
\begin{align}
&\langle\ \varphi \ |\ \psi\ \rangle =
e^{{1\over 4i\ga}(2\varphi\psi-\varphi ^2)},\label{x}\\
&\lambda (\zeta\ |\ \psi )=
e^{-{1\over 2i\ga}\zeta \psi}
\ \Phi (\psi -\zeta)\ e^{{\pi +\ga\over\ga}(\psi-\zeta)},\non
\end{align}
and the function $\Phi (\varphi )$ satisfies the functional equation:
\begin{align}
&\frac{\Phi (\varphi +i\ga)}{\Phi (\varphi -i\ga)}={1\over 1+e^\varphi}
\label{phieq}
\end{align}
The solution to this equation is
$$\Phi (\varphi)=\exp \(\ \int\limits _{\mathbb{R}+i0}{e^{ik\varphi}\over
4\sinh \ga k\ \sinh \pi k}{dk\over k}\)$$
This wonderful function and its applications can be found in \cite{f}.

As usual we want to consider 
$Q(\varphi _{1},\cdots ,\varphi _{2g+2}\ |\ \zeta \ |\ \psi _1,\cdots ,\psi _{2g+2})$ 
as the kernel of an operator:
$$Q(\varphi _{1},\cdots ,\varphi _{2g+2}\ |\ \zeta \ |\ \psi _1,\cdots ,\psi _{2g+2})=
\langle\  \varphi _{1},\cdots ,\varphi _{2g+2}\ |\ \qqq (\zeta)
\ |\ \psi _1,\cdots ,\psi _{2g+2}\ \rangle $$
The subtle point is that we have to use
mixed representations: the vectors $|\ \psi\ \rangle $ are the eigenvectors of 
the operators
$$w\equiv e^{\psi }=uvu $$
Notice that this justifies the
notation $\langle\ \varphi \ |\ \psi \ \rangle $ in (\ref{x}), and that
$$[\ \psi ,\varphi \ ]=2i\ga $$
The operators $\qqq (\zeta)$ satisfy the equations
\begin{align}
&
(-1)^{g+1}\tt (z)\qqq (\zeta)=\qqq (\zeta +i\ga)+\qqq (\zeta -i\ga)\label{tq}
\end{align}
This is famous Baxter's equation.


Before going further let us discuss the properties of the operator $\qqq (\zeta)$.
We have
\begin{align}
&\overline{\Phi(\varphi )}=\Phi (\bar{\varphi}); \non\\& \Phi (\varphi )\sim
\exp \({1\over4i\ga}\varphi ^2\),\qquad \text{as}\quad
 \varphi \to\infty\label{bar}
\end{align}
so, the kernel of $\qqq (\zeta)$ for $\zeta \in \mathbb{R}$ is
an oscilating function, and it is rather clear that our operator
is well defined on the functions of $\psi _j$ of Schwartz class
($\text{Sch}_{\psi}$) sending them to functions of $\varphi _j$
which are also of Schwartz class
($\text{Sch}_{\varphi}$). Using the equations (\ref{bar}) one easily finds
the kernel $\langle\  \psi \ |\ \qqq ^*(\zeta)\ |\ \varphi \ \rangle $
of the adjoint operator $\qqq ^*(\zeta)$
(we consider the case of real $\zeta$). Further, notice that the 
l-operator can
be rewritten as
\begin{align}
l(z)=
{1\over\sqrt{z}}\begin{pmatrix}z u&
-qu^{-1} w\\
zuw^{-1} &0\end{pmatrix}
\end{align}
Applying to this l-operator the same procedure as before one finds that
$\qqq ^*(\zeta )$ also solves the Baxter equation (\ref{tq}):
$$(-1)^{g+1}\tt (z)\qqq ^*(\zeta )=\qqq ^*(\zeta +i\ga )
+\qqq ^*(\zeta -i\ga)$$
It can be shown that actually 
$$\qqq (\zeta )=\qqq ^*(\zeta )\qquad \text{for}\qquad \zeta\in\mathbb{R}$$
Considering the kernel of the
operator $\qqq ^*(\zeta )$
one finds that this operator acts from
$\text{Sch}_{\varphi}$ to $\text{Sch}_{\psi}$. So, the products 
$\qqq (\zeta )\qqq (\zeta ' )$ are well defined at least for 
$\zeta ,\zeta ' \in\mathbb{R}$.

We want to show that the operators $\qqq (\zeta )$ constitute a commutative
family:
\begin{align}
&[\ \qqq (\zeta ),\qqq (\zeta ')\ ]=0
\label{commu}
\end{align}
To this end we want to show that the operator $\qqq (\zeta )$
can be rewritten as
\begin{align}
&\qqq (\zeta )=\text{tr}_a\ 
\bigl(\ll _{a 2q+2}(\zeta)\ \cdots \ \ll _{a 1}(\zeta)\bigr)
\label{q=lll}
\end{align}
where the operators $\ll _{a j}(\zeta)$ act in the tensor
product of the ``auxilary space'' labeled by $a$ and of the 
``quantum space''
where $\varphi _j$, $\psi _j$ act. 
Actually in our case the ``auxilary space'' will be isomorphic
to the ``quantum space'', i.e. we shall have a universal l-operator.
If the operators $\ll _{aj}(\zeta)$
satisfy Yang-Baxter equations with some R-matrix then the commutativity
(\ref{commu}) follows from the standart argument.

To find the representation (\ref{q=lll}) rewrite (\ref{ker}) as 
\begin{align}
&\qqq (\zeta )=e^{{1\over 2}(1+{\pi\over\ga})\zeta+{1\over 4i\ga}\zeta ^2 }
\int
\prod _{j=1}^{2g+2}d\varphi _j'd\psi _j'
\langle\ \psi _j'\ |\ \ll _{a j}(\zeta)\ |\ \varphi _j'\ \rangle
\langle\ \varphi _j' \ |\ \psi _{j-1}'\ \rangle 
\non
\end{align}
where $\varphi _j'$, $\psi _j'$ are operators acting in the
``auxiliary space'',
$\psi _{0}'=\psi _{2g+2}'$. So, (\ref{q=lll}) indeed takes
place  
if the 
kernel of the ``universal'' l-operator is given by
$$\langle\  \varphi '\ |\ \otimes\langle\  \psi\ |\ 
\ \ll (\zeta)\ \ |\ \psi '\ \rangle\otimes \ |\ \varphi \ \rangle =
\delta (\varphi -\varphi ')\delta (\psi-\psi ')
\lambda (\zeta\ |\ \varphi-\psi )$$
Hence the formula (\ref{q=lll}) holds
for the operators $\ll _{aj}(\zeta)$ of the form:
$$\ll _{12}(\zeta) =\mathcal{P}_{12}\lll _{12}(\zeta )$$
where $\mathcal{P}_{12}$ is the operator of permutation, and the
operator $\lll _{12}(\zeta )$ acts in the tensor product as follows:
$$ \lll _{12}(\zeta )=
\lambda (\zeta |\ \varphi \otimes I-I\otimes \psi  )$$
Thus the operator $\qqq (\zeta)$ can be considered
as trace of ``universal'' monodromy matrix and the 
commutativity (\ref{commu}) follows from the Yang-Baxter equation:
\def\rrr{\widehat{\mathcal{R}}}
\begin{align}
\rrr _{12}(\zeta _1 -\zeta _2 )\ \lll _{23}(\zeta _1)
\ \lll _{12}(\zeta _2)=
\lll _{23}(\zeta _2)\ \lll _{12}(\zeta _1)\ \rrr _{23}(\zeta _1 -\zeta _2 )
\label{yb}
\end{align}
with the simple r-matrix:
$$\rrr _{12}(\zeta )=
\exp \({(I\otimes \psi -\varphi \otimes I)\zeta\over 2i\ga }\)$$
The Yang-Baxter equation (\ref{yb}) in our case is almost trivial.
In the case of the more general l-operator
$\widehat{l}(z,\kappa )$ mentioned above we would need to use
a more complicated r-matrix and the proof of Yang-Baxter equations
needs some non-trivial identities \cite{fkv}.

The self-ajoint
(for real $\zeta _{1},\zeta _2$) 
operators $\qqq (\zeta _1)$, $\qqq (\zeta _2)$ commute,
hence the eigen-vectors of $\qqq (\zeta )$ do not depend on $\zeta$.
Actually, the operator $\qqq (\zeta )$ is an entire function of $\zeta$.
The kernel of $\qqq (\zeta )$ has poles, but in the process of analytical
continuation the poles never pinch the contour of integration.
The Baxter's equation (\ref{tq}) implies that $\qqq (\zeta _1)$ 
and $\tt (z _2)$ 
also commute.
Suppose that $\qq (\zeta )$ ant $t(z)$ are eigen-values
of these operators, 
due to the equation  (\ref{tq})
they satisfy 
\begin{align}
&(-1)^{g+1}t(z)\qq (\zeta)=\qq (\zeta +i\gamma)+\qq (\zeta -i\gamma)
\label{tq1}
\end{align}

Let us discuss further analytical properties of $\qq(\zeta)$.
Since the operator $\qqq (\zeta )$ is an 
entire
functions of $\zeta$ the eigen-value
$\qq(\zeta)$ is an  entire function as well.
As it has been said $\tt (0)=\tt _{g+1}$ belongs to the center of the algebra
defined by the commutation relations (\ref{rll}), so, we can fix it.
It is convenient to put $\tt _{g+1}=(-1)^{g+1}2$ which allows to reqiure
that 
\begin{align}
\qq (\zeta)\to 1, \qquad \zeta\to -\infty
\label{to0}
\end{align}
From quasi-classical consideration which are completely parallel to
those from \cite{gp,toda} it is naturally to conjecture that the
eigenvalues of $\qq (\zeta )$ have zeros only on the real axis
and that asymptotically for $\zeta\to\infty$ one has:
\begin{align}
&\qq (\zeta)\sim 
e^{-(g+1)(1+{\pi\over\ga})\zeta}
\cos \({(g+1)\zeta ^2\over \ga}+{\pi\over 4}\)
\label{asy}
\end{align}

The important question is whether the equations (\ref{tq1})
together with the analytical properties described above
are sufficient to find the spectrum of commuting
Hamiltonians.
In our opinion it is impossible, the additional
information is needed which is provided in the following
section.

\section{Duality.}

Consider the function $\Phi(\zeta )$. The most interesting
property of this function is its
duality:
together with the equation (\ref{phieq}) it satisfies the
equation
\begin{align}
&\frac{\Phi (\varphi +i\pi)}{\Phi (\varphi -i\pi)}
={1\over 1+e^{{\pi\over\ga}\varphi}}
\non
\end{align}
Using this property and the definition of the
operator $\qqq (\zeta )$ one finds that there is
dual equation for $\qqq (\zeta )$:
\begin{align}
(-1)^{g+1}\TT (Z)\qqq (\zeta )=\qqq (\zeta +\pi i)+\qqq (\zeta -\pi i)
\label{Tq}
\end{align}
where
$$Z=e^{{2\pi\over\ga}\zeta},$$
and
$\TT (Z)$ is the trace of the monodromy matrix 
$$\MT (Z)=L_{2g+2}(Z)\ \cdots \ L_1(Z)$$
with
\begin{align}
L(Z)={1\over\sqrt{Z}}
\begin{pmatrix}Z U^{-1}&
-QVU\\
ZV^{-1}U^{-1} &0\end{pmatrix}\non
\end{align}
The dual operators 
$$U=e^{{\pi\over\ga}\varphi },
\qquad V=e^{\pi i{d\over d\varphi }}$$ 
satisfy the commutation
relations 
$$UV=QVU$$
with dual
$$Q=e^{i{\pi ^2\over \ga}}$$
The only non-trivial commutation relations of
$u,v$ with $U,V$
are 
$$uV=-Vu,\qquad vU=-Uv$$
which means that
$$S\ (l(z )\otimes I)(I\otimes L(Z))=(I\otimes L(Z))(l(z )\otimes I) \ S$$
with
$S=\sigma ^3\otimes\sigma ^3$. From here it is obvious that
$$[\ \tt (z),\TT (Z)\ ]=0$$

All that is the result of manifest duality of the kernel
of $\qqq (\zeta)$ with respect to change:
\begin{align}
\ga\to{\pi ^2\over \ga},
\qquad \zeta\to{\pi\over\ga}\zeta,\qquad \varphi _j\to{\pi\over\ga}\varphi _j,
\qquad \psi _j\to{\pi\over\ga}\psi _j\non
\end{align}

It is clear that $\TT (Z_1)$ and $\qqq (\zeta _2 )$
commute, so, the equation (\ref{Tq}) implies 
the equation for eigen-values:
\begin{align}
(-1)^{g+1}T (Z)\qq (\zeta )=\qq (\zeta +\pi i)+\qq (\zeta -\pi i)
\label{Tq1}
\end{align}

The function $\qq (\zeta )$ is not an entire function
of $z$ as it is the case in other situations 
(for example \cite{blz}), that is why the equation (\ref{tq})
alone does not look strong enough to define it.
However, the equation (\ref{Tq}) controlling the 
behaviour of $\qq (\zeta )$ under the $2\pi i$ -rotation
in $z $-plane must provide the missing information.

So, our
main conjecture is the following
\vskip0.3cm
\noindent
{\bf Conjecture 4.} {\it
The spectrum on $\tt (z)$ (and, simultaneously, of $\TT (Z)$)
is described by all solutions of the equations (\ref{tq1})
and (\ref{Tq1}) such that }
\newline
1. {\it $t(z)$ and $T(Z)$ are polynomials of degree 
$g+1$.}\newline
2. {\it $\qq (\zeta )$ is an entire function of $\zeta$.}\newline
3. {\it $\qq (\zeta )$
satisfies (\ref{to0}) and (\ref{asy}).}\newline
4. {\it All the zeros of $\qq (\zeta )$ 
in the strip} 
$-(\pi +\gamma)<
\text{Im}(\zeta )<(\pi +\gamma)$ 
{\it are real.}


\section{Separation of variables.}

The relation of integrable models to the algebraic geometry
can be completely understood in the framework of
separation of variables. 

We have already mentioned that
$$[\ \bb (z),\bb (z ')\ ]=0$$
which implies commutativity of the operators $\zz _j$
defined as roots of $\bb (z)$.
Consider the operators 
$$\ww _j=(-1)^{g+1}q\ \dd (\overleftarrow{\zz}_j)
$$
where $d(\overleftarrow{\zz _j})$ means that $\zz _j$ which does not
commute with coefficients of $\dd(z)$ is substituted to this
polynomial form the left.
Following Sklyanin \cite{skl} one shows that
$$\zz _j\ww _k=\ww _k\zz _j, \quad j\neq k ;\qquad \zz _j\ww _j
=q^2\ww _j\zz _j $$
and
\begin{align}
&\ww _j^2-\ww _j
\tt (
\overleftarrow
{\zz}_j)+1=0
\label{sep}
\end{align}
Introduce the
operators
$$\zzeta _j=\textstyle{\frac 1 2}\log (\zz _j )$$
and consider the wave-function corresponding to given
set of eigen-values of integral of motion $\tt _1,\cdots ,\tt _g$
in $\zzeta $-representation.
The equation (\ref{sep}) implies \cite{skl} that we can look
for  this wave function in the form:
$$\langle \ \zeta _1,\cdots ,\zeta _g\ |\ t_1,\cdots ,t_g\ \rangle
=\qq (\zeta _1)\cdots \qq(\zeta _g )$$
where $\qq (\zeta )$ satisfies the equation:
$$\qq (\zeta +i\ga )+\qq (\zeta -i\ga )=(-1)^{g+1}t(z)\qq (\zeta )$$
where $t(z)$ is constructed from the eigen-values $t$.
This equation coincides with the equation (\ref{tq}) written
for particular eigen-values. So, 
following \cite{gp} we claim that the wave function
in separated variables is defined by eigen-value of the
operator $\qqq $ which connects two different approaches
to integrable models. 

\def\F{\mathcal{F}}
Notice that the vector $|\ t_1,\cdots ,t_g\ \rangle$
is eigen-vector for the operators $\TT _1,\cdots ,\TT _g$ since
the function $\qq (\zeta )$ satisfies
the equation (\ref{Tq}). 
In order to identify explicitly the eigen-vales
of $\TT _1,\cdots ,\TT _g$
we shall write 
$|\ t_1,\cdots ,t_g;T_1,\cdots ,T_g\ \rangle$.

We have the algebra of operators $\A (q)$ and the dual
algebra $\A (Q)$ which act in the same space
$\mathfrak{H}$. All the operators from   $\A (q)$
commute with the operators from $\A (Q)$.
The fundamental property of $\A(q)$
is that it is spanned as linear space by
the elements of the form (\ref{f=PHP})
according to Conjecture 3. Similar fact must be true
for $\A (Q)$. Taking these facts together one
realizes the algebra $\A (q)\cdot\A (Q)$
is spanned by the elements of the form:
\begin{align}
\X=&xX=p_L(\tt _1,\cdots ,\tt _g)P_L(\TT _1,\cdots ,\TT _g)\\
&\times
g(\bb _1,\cdots ,\bb _g)G(\BB _1,\cdots ,\BB _g)\ P_R(\TT _1,\cdots ,\TT _g)
p_R(\tt _1,\cdots ,\tt _g)
\non
\end{align}
We denote by 
$h(\zz _1,\cdots ,\zz _g)$ and $H( \ZZ _1,\cdots ,\ZZ _g)$
anti-symmetric polynomials 
obtained from $g(\bb _1,\cdots ,\bb _g)$ and $G(\BB _1,\cdots ,\BB _g)$,
for example:
\begin{align}
h(\zz _1,\cdots ,\zz _g)=\prod \zz _i\prod\limits _{i<j}(\zz _i-\zz _j)
g(\bb _1(\zz _1,\cdots ,\zz _g),
\cdots ,\bb _g(\zz _1,\cdots ,\zz _g))
\non
\end{align}


Let us consider the matrix element of the
operator $\X$ between two eigen-vectors of Hamiltonians.
The wave functions are real for real $\zeta $.
By requirement of self-ajointness of $\tt (z)$ and $\TT (Z)$
one defines the scalar product ({\it cf.} \cite{skl}).
The matrix element in question is
\begin{align}
&\langle \ t_1,\cdots t_g\ ;\ T_1,\cdots ,T_g\ |
\ \X \ 
|\ t'_1,\cdots t'_g\ ;\ T'_1,\cdots ,T'_g\ \rangle =\non\\&=
p_L(t_1,\cdots ,t_g)p_R(t'_1,\cdots ,t'_g)
P_L(T_1,\cdots ,T_g)P_R(T'_1,\cdots ,T'_g)
\label{me}\\&\times
\int\limits _{-\infty}^{\infty}d\zeta _1\cdots
\int\limits _{-\infty}^{\infty}d\zeta _g
\ h(z_1,\cdots ,z_g)H(Z_1,\cdots ,Z_g )
\prod\limits _{j=1}^g \qq (\zeta _j)\qq ' (\zeta _j)
\non
\end{align}

When does the integral for the
matrix element (\ref{me}) converge?
Suppose that 
$$h\sim z_j^{g+k+1},\ H\sim Z_j^{g+l+1} 
\qquad\text{when}\quad \zeta _j\to\infty$$
then the integrand in the matrix element behaves when $\zeta _j\to\infty$
as
$$\exp 2\zeta _j\bigl((k-1)+{\pi\over\gamma}(l-1)\bigr)$$
hence for generic $\gamma $ the integral converges only
if $k=1,l=0$ or $k=0,l=1$. When $\gamma $ is small
we can allow the operators with $l=0$ and $k<{\pi\over\gamma}$,
oppositely, when $\gamma$ is big the operators with $k=0$
and $l<{\gamma\over\pi }$ are allowed. The limits $\gamma\to 0$
and $\gamma\to\infty$ are two dual quasi-classical limits.
For these limits the operators $l=0,\forall k$ and $k=0,\forall l$ 
respectively
define the classical observables.
At least these operators must be defined
in the quantum case: if the quantization procedure
makes sense the principle of correspondence must hold.
Hence, the fact that in general only two operators 
with $k=1,l=0$ or $k=0,l=1$ 
lead to convergent integrals 
means that some regularization of these integrals is needed.
The regularized integrals  in question must allow to define
the matrix element (\ref{me}) for arbitrary $k, l$, they have to
coincide with usual integrals whenever the latter are applicable,
they must satisfy some additional requirements which
will be discussed in the Section 9.
The origin of these additional requirements
is in the cohomological construction explained in Section 4.

Notice that any anti-symmetric with respect to 
$z_1,\cdots ,z_g$ and $Z_1,\cdots ,Z_g $ polynomial
$$
h(z_1,\cdots ,z_g )H(Z_1,\cdots ,Z_g)
$$
can be presented as linear combination 
of products of Schur-type
determinants
$$\text{det}|z_i^{k_j}|\ \text{det}|Z_i^{l_j}|$$
where $\{k_1,\cdots, k_g\}$ and $\{l_1,\cdots, l_g\}$ 
are arbitrary sets of positive integers.
So, the integrals (\ref{me}) can be expressed in terms
of 1-fold integrals
\begin{align}
&\langle \ l\ |\ L\ \rangle\simeq
\int\limits _{-\infty}^{\infty}
\qq (\zeta )\qq '(\zeta )l(z)L(Z)d\zeta
\label{lL}
\end{align}
where $l$ and $L$ are polynomials
such that $l(0)=0$ and $L(0)=0$. 
The symbol $\simeq $
means that the integrals in RHS are not always defined,
the regularization is defined in th Appendix B,
in the next section we describe results of this regularization..

\section{Deformed Abelian differentials.}

In Appendix B we define the polynomials $s_k(z)$.
These polynomials are of the form
\begin{align}
& s_{k}(z)=z^{g+1+k}, \qquad -g\le k\le 0\non\\
& s_{k}(z)=\frac {1}{i\gamma}
\(\frac {q^k-1}{q^k+1} \)
\ z^{g+1+k}+\cdots,
\quad k\ge 1\label{s_k}
\end{align}
where $\cdots $ stands for terms of lower degree
(containing $t_j$, $t_j'$ in coefficients)
explicitly given in Appendix B.

In the classical case every 
polynomial 
defines  an Abelian differentials on the
affine curve $X-{\infty ^{\pm}}$.
Similarly we consider
the polynomials $s_k$    
as corresponding to ``deformed Abelian
differentials.'' Let us be more precise.
The regularized integrals
are defined in Appendix B in such a way that
they satisfy several conditions. First of them is
\begin{align}
\langle \ s_k\ |\ S_l\ \rangle =0\qquad
k\ge g+1,\quad\forall l\label{s}\\
\langle \ s_k\ |\ S_l\ \rangle =0\qquad
\forall k,\quad l\ge g+1\label{S}
\end{align}
Due to (\ref{s}) we consider the polynomials $s_k$,
$k\ge g+1$ as corresponding to exact forms.
The polynomials $s_k$ with $k=-1,\cdots ,-g$ correspond
to first kind differentials, 
$s_{o} $
corresponds to
the third kind one, $s_k$ with $ k=1,\cdots ,g$
correspond to second kind differentials. 

Explicitly the relation with classical case is as
follows. Consider the case $t(z)=t'(z)$ and take the limit:
$$r_k=z^{-1}\lim _{\gamma\to 0}s_k(z)$$
Then the classical Abelian differential related to $s_k$
is 
$$\mu _k=\frac {r_k(z)}{y}dz$$

Similar interpretation can be given to $S_k$
which correspond to Abelian differentials in dual
classical limit $\gamma \to\infty$. However, the 
most interesting feature of our construction is
that 
together with this cohomological
interpretation an alternative ``homological'' one is possible.
The polynomials $S_k$, $k\ge g+1$ correspond to
retractable cycles according to (\ref{S}).
The polynomials $S_k$ for $k=\pm 1,\cdots ,\pm g$
are interpreted as analogues $a$ and $b$ cycles
$\delta _k$ on the ``deformed affine
curve'', $S_{0}$ corresponds to cycle $\delta _0$
around $\infty ^+$
which is
non-trivial on the affine curve.
The pairing $\langle\ l\ |\ L\ \rangle $ defines
the integral of differential defined by $l$ over
cycle defined by $L$. 
The asymptotic of the integrals $\langle\ l\ |\ L\ \rangle $
in the classical limit $\gamma \to 0$
are, indeed, described by Abelian integrals.
Certainly, the opposite interpretation
($l$-cycle, $L$-differential )
is possible which corresponds to dual classical
limit. It is not the first time that this kind
of objects appears \cite{abel}, but it is the
first time that we observe real duality
between two classical limits.

Let us define the pairing between two polynomials
$l_1$ and $l_2$ 
\begin{align}
&l_1\circ l_2=\lim _{\Lambda\to\infty}
\ \int\limits _{\Lambda} ^{\Lambda +i\pi }
 \bigl[\qq (\zeta )\qq '(\zeta )l_1(z)\delta ^{-1}_{\gamma}
(\qq\qq 'l_2)(\zeta -i\pi )
+\non\\ 
&+\qq (\zeta -\pi i)\qq '(\zeta-i\pi  )
l_1(z)\delta ^{-1}_{\gamma}
(\qq\qq 'l_2)(\zeta -i\gamma)\bigr]d\zeta
\label{par}
\end{align}
One can show that these formulae give
well-defined anti-symmetric pairings
which correspond classically to
natural pairing between meromorphic differentials
$$\omega _1\circ\omega _2=\text{res}_{p=\infty ^+}
\(\omega _1(p)\int\limits ^p\omega _2 \)$$
The polynomials $s_{\pm j}$ and 
for $j=1,\cdots ,g$
constitute canonical basis:
\begin{align}
\quad s_k\circ s_l=\text{sgn}(k-l)\delta _{k, -l}
\non
\end{align}
Similarly, to introduce the definition of $L_1\circ L_2$ it
is sufficient to do necessary replacements in (\ref{par}):
$l_i\leftrightarrow L_i$, $z\leftrightarrow Z$,
$\gamma\leftrightarrow \pi$. The polynomials $S_{\pm j}$
are canonically conjugated.
 
The following anti-symmetric polynomials play role
of 2-forms $\sigma$ and $\sigma '$ used in classics:
\begin{align}
&c(z_1,z_2)=
\sum\limits _{j=1}^g\(s_{-j}(z_1)s_j(z_2)-s_j(z_1)s_{-j}(z_2)\),
\label{c}\\
&C(Z_1,Z_2)=
\sum\limits _{j=1}^g\(S_{-j}(Z_1)S_j(Z_2)-S_j(Z_1)S_{-j}(Z_2)\)
\non
\end{align}

As usual \cite{abel} the most important property
of deformed Abelian integrals is that the Riemann
bilinear relations remain valid after the deformation.
Namely, consider the the following $2g\times 2g$
period matrix:
$P$ with the matrix elements
\begin{align}
P_{kl}=&\langle\ s_k\ |\ S_l\ \rangle
&k,l=-g,\cdots , -1,1,\cdots ,g\non
\end{align}
The deformed Riemann bilinear identity is formulated as
\vskip 0.3cm
\noindent
{\bf Proposition 2.} {\it The matrix $P$ belongs to the symplectic
group:}
\begin{align}
&P\in Sp(2g) \label{riemann}
\end{align}
\vskip 0.3cm
\noindent
This Proposition 2 is equivalent to a number of
bilinear relations between the deformed Abelian
integrals. Proving them 
it is convenient to consider the domain of small
$\gamma$ 
($\gamma <\pi/n$)
when the regularization of integrals simplifies,
and then continue analytically with respect to $\gamma$.
Still the proof  is rather
complicated technically: it is based on
non-trivial properties of the regularized
integrals. We do not give this bulky proof
here.

There is one more relation for deformed Abelian integrals.
One can check that
\begin{align}
&\langle \ d\ |\ S_k\ \rangle=0,\quad\forall k,\qquad
d=\sum\limits _{j=1}^g (t_j-t_j')s_{-j}\label{d=0}\\
&\langle \ s_k\ |\ D\ \rangle=0,\quad \forall k,\qquad
D=\sum\limits _{j=1}^g (T_j-T_j')S_{-j} \label{D=0}
\end{align}
The relations (\ref{d=0}) do not have direct analogue
in terms of Abelian integrals, recall that
we put $t_j=t_j'$ taking classical limit which turns
the relation into triviality.
However, there is another way of taking the classical limit
where this equation is important \cite{toda}.

\section{Return to quantization of affine Jacobian.}

Let us return to the main subject of this paper:
quantization of affine Jacobian. 
Consider any $x\in \A (q)$.
Such $x$ is identified with $x\cdot I\in \A(q)\cdot \A(Q)$,
so, due to Conjecture 3  the matrix element of $x$
can be presented as
\begin{align}
&\langle \ t_1,\cdots t_g\ 
|
\ x \ 
|\ t'_1,\cdots t'_g\ 
\rangle =
p_L(t_1,\cdots ,t_g)p_R(t'_1,\cdots ,t'_g)
\label{mef}\\&\times
\int\limits _{-\infty}^{\infty}d\zeta _1\cdots
\int\limits _{-\infty}^{\infty}d\zeta _g
\ h(z_1,\cdots ,z_g)
H_I(Z_1,\cdots ,Z_g )
\prod\limits _{j=1}^g \qq (\zeta _j)\qq ' (\zeta _j)
\non
\end{align}
where the eigen-values of $\TT _j$ are the same at the 
left and at the right, so, we do not write them explicitly,
the polynomial $H_I$ which corresponds to $X=I$ is given by
$$H_I(Z_1,\cdots ,Z_g )=\prod\limits _{j=1}^g Z_j
\prod\limits _{i<j}(Z_i-Z_j)
$$
notice that 
\begin{align}
H_I=S_{-1}\wedge\cdots\wedge S_{-g}\label{H=sss}
\end{align}

The formula for the matrix elements (\ref{mef})
for small $\gamma$ (when no regularization of integrals is needed)
can be deduced rigorously starting from realization
of $\A (q)$ defined in Section 4. The following 
equations follow respectively 
from (\ref{s}), (\ref{d=0}), (\ref{riemann})
(recall notations of Section 4):
\begin{align}
&s_k\wedge \V ^{g-1}\simeq 0,\quad k\ge g+1\label{eq1}\\
&c\wedge \V ^{g-2}\simeq 0, \label{eq2}\\
&d\wedge \V  ^{g-1}\simeq 0,\label{eq3}
\end{align}
where $\simeq $ means that these expression vanish being substitute
into the integral (\ref{mef}).
The equation (\ref{eq2}) needs explanation.
To prove this equation one has to take into account
the Riemann bilinear identity (\ref{riemann})
and the formula (\ref{H=sss}); notice that 
$$S_{-i}\circ S_{-j}=0,\quad 1\le i,j\le g$$

The formula for the matrix elements (\ref{mef})
can be rigorously deduced for small $\gamma$. Hence
the equations (\ref{eq1}), (\ref{eq2}), (\ref{eq3})
lead to equation for operators. The latter
equations are obtained applying  the operation $\chi $
(Section 4):
\begin{align}
&\chi \bigl(s_k \wedge \V ^{g-1}\bigr)=0,\quad k\ge g+1      \label{i1}\\
&\chi \bigl( c\wedge \V ^{g-2}\bigr)=0,         \label{i2}\\
&\chi \bigl(d\wedge\V ^{k-1}\bigr)=0           \label{i3}
\end{align}
We conclude that the 
formulae for the polynomials $s_k$ needed in  Section 4
are exactly the same as given in (\ref{s_k}).
Thus we put together the algebraic part of
this work
with the analytical one.

On the other hand the equations (\ref{i1}, \ref{i2}, \ref{i3})
are of purely
algebraic character, so, if they are
valid for small $\gamma$ they must be valid always.
That is why we regularized the integrals for
matrix element in order that the equations 
(\ref{eq1}), (\ref{eq2}), (\ref{eq3}) hold for any $\gamma$.

Moreover, there is a dual model and
we can consider the operators $\X=xX$
from $\A(q)\cdot\A (Q)$. The equations 
(\ref{i1}), (\ref{i2}), (\ref{i3}) 
and dual equations still have to be valid.
The regularized integrals are defined in such a way that
it is the case. The equations (\ref{i1}), (\ref{i3})
and their duals clearly follow from (\ref{s}), (\ref{S})
and (\ref{d=0}),  (\ref{D=0}).
The most interesting is the equation (\ref{i2}).
Due to the Riemann bilinear relation
this equation follows from 
\begin{align}
c\wedge \V ^{g-2}\simeq 0
\label{cV}
\end{align}
which is true if the sub-space $c\wedge \V ^{g-2}$
of the space $\V ^{g}$
is convoluted with the sub-space 
$$\frac{\V _g}{C\wedge \V _{g-2}}$$
where $\V _k$ is the same as $\V ^k$ for dual model
(this notation is not occasional: the space $\V _k$
plays role of $k$-cycles for $k$-forms from $\V ^k$).
In other words we impose the equation 
\begin{align}
C\wedge \V _ {g-2}=0
\non
\end{align}
and the dual equation
(\ref{cV})
is imposed automatically due to the Riemann bilinear relation.
 
Let us discuss the classical limit in some
more details. Consider the hyper-elliptic curve $X$.
If we realize this curve as characteristic equation of
classical analogue of the monodromy
matrix $\mt (z)$ (\ref{mt}) the branch points of the curve
can be shown real non-negative. Actually requiring $t_{g+1}=(-1)^{g+1}2$
we put one of branch points at $z=0$. Thus the branch points are
$0=q_1$ $<$ $\cdots $ $<$ $q_{2g+2}$.
The Riemann surface is realized as two-sheet covering
of the plane of $z$ with cuts $I_k=[q_{2k-1},q_{2k}]$,
$k=1,\cdots ,g+1$. The canonical a-cycles $\delta _{-j}$
and b-cycles $\delta _j$ are shown on the {\it fig. 1}
follows:
\vskip 1cm
\hskip -0.5cm  
\epsffile{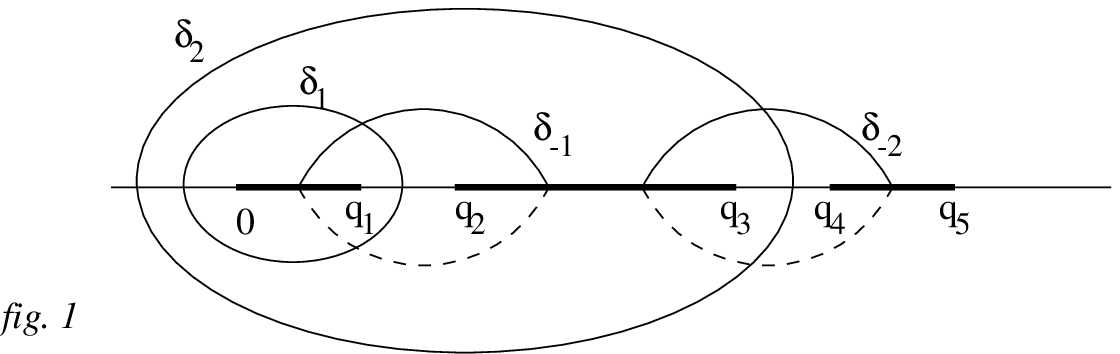}
\vskip 0.2cm
\noindent
Under the classical dynamics every of 
the separated variables $z_j$ oscillates in the interval:
$q_{2j-1}\le z_j\le q_{2j}$, topologically 
it corresponds to motion along the a-cycle $\delta _{-j}$.
One can show that the integral $\langle \ s_k\ |\ S_{-j}\ \rangle$
is described in the classical limit $\gamma \to 0$
by $\delta _{-j}$ of differential $\mu _k$. Thus the $g$-cycle
(\ref{H=sss}) corresponds to classical trajectory
$\delta _{-1}\wedge\cdots\wedge\delta _{-g}$.
Recall that the cycle (\ref{H=sss}) corresponds
to insertion of unit operator of the dual model.
Introducing other dual operators one gets
integrals with respect to both a-cycles and b-cycles.
Classically the coresponding trajectories are 
not real, but
the factorization by $\sigma '\wedge W_{m-2}$
in (\ref{H_m}) guarantees that the classical non-real 
trajectories are not singular.
We would like to finish this paper with this topological
interpretation of the dual model.

\section{Appendix A.}

In this Appendix we shall give the
canonical definition of affine Jacobi variety
$\Ja$.
Consider hyper-elliptic curve $X$ of genus $g$:
\begin{align}
&w^2 -t(z)w+1
=0\non
\end{align}
We have the canonical basis with
a-cycles $\delta _k$, $-g\le k<0$ and b-cycles
$\delta _k$, $0< k\le g$.
Associate with this basis
the basis of normalized holomorphic differentials
$\omega _j$:
$$\int\limits _{\delta _{-i}}\omega _j=\delta _{ij},
\quad B_{ij}=\int\limits _{\delta _i}\omega _j$$
The Jacobi variety of this curve is the $g$-dimensional
complex torus:
$$
\J=\frac{\mathbb{C}^g}{\mathbb{Z}^g\times B\mathbb{Z}^g}
$$
With every point $p\in X$ we identify the point
$\alpha (p)\in \J$ with coordinates:
$$\alpha _j(p)=\int\limits _{b}^{p}\omega _j$$
for the reference point
$b$ it is convenient to take
one  of the branch points.
The curve $X$ has two points over the
point $z=\infty$, denote them by $\infty ^{\pm}$
and consider the $(g-1)$-dimensional subvariety
of $\J$ defined by 
$$\Theta ^{\pm}=\{\zeta\in\J\ |\ \theta (\zeta+\alpha (\infty ^-))
\theta (\zeta+\alpha (\infty ^+))=0\}$$
where $\theta$ is Riemann theta-function.
It can be shown that there exist an isomorphism:
\begin{align}
\Ja\simeq\J-\Theta ^{\pm}
\label{isom}
\end{align}

The equivalence of this description  with the
description in terms of divisors (Section 1)
is due to the Abel map
$X[g]\to J(t)$ 
explicitely given by
$$\zeta =\alpha (\P)+\Delta,\quad \alpha (\P)=\sum \alpha ( p_j)$$
where $\Delta $ is the Riemann characteristic.

\section{Appendix B.}

In this Appendix we describe the regularization
of integrals which has been used in the paper.

Define:
\begin{align}
&\delta _{\xi }(f(\zeta))=f(\zeta +i\xi )-f(\zeta),\non\\
&\Delta _{\xi} (f(\zeta))=f(\zeta +i\xi )-f(\zeta -i\xi )\non
\end{align}
Introduce the polynomials 
\begin{align}
&s_k(z)=\frac 1 {2i\gamma}\bigl\{ 
t(z)\Delta _{\gamma} ^{-1}[z^{k-g-1}t(z)]_>+
t'(z)\Delta_{\gamma} ^{-1}[z^{k-g-1}t'(z)]_>-\non\\
&\qquad\qquad -t(z)\Delta _{\gamma}^{-1}
[z^{k-g-1}q^{2(g+1-k)}t'(zq^{-2})]_>-\non\\
&\qquad\qquad -t'(z)\Delta _{\gamma}^{-1}
[z^{k-g-1}q^{2(g+1-k)}t(zq^{-2})]_>-\non\\
&\qquad\qquad -\textstyle{{1\over 2}}\(t'(z)[z^{k-g-1}t(z)]_> 
+ t(z)[z^{k-g-1}t'(z)]_>\)+\non\\
&\qquad\qquad +(q^{2(g+1-k)k}-q^{2(k-g-1)})[z^{k-g-1}]_>\bigr\},\qquad 
k\ge 0;\non\\
&s_k(z)=z^{g+1+k},\qquad\qquad
-g\le k\le 0;\non
\end{align}
where the notation $[\quad ]_>$ means that only the positive
degrees of Laurent series in brackets are taken.
Obviously $\text{deg}(s_k)=g+1+k$.
Further, with every function $f(\zeta)$
associate the functions:
\begin{align}
u[f](\zeta)=\frac 1 {2i\gamma}&\bigl\{
            t(z)\Delta_{\gamma} ^{-1}\(f(\zeta)t(z)\)+
            t'(z)\Delta_{\gamma} ^{-1}\(f(\zeta)t'(z)\)-\non\\
            &-t(z)\Delta _{\gamma}^{-1}\(f(\zeta -i\gamma)t'(zq^{-2})\)-
            t'(z)\Delta _{\gamma}^{-1}\(f(\zeta -i\gamma)t(zq^{-2})\)-\non\\
            &-f(\zeta)t(z)t'(z)
            +f(\zeta +i\gamma )-f(\zeta -i\gamma )\bigr\},\non\\
v[f](\zeta)=\frac 1 {2i\gamma}&\bigl\{
             (-1)^{g+1}\bigl(\Delta _{\gamma}^{-1}\(f(\zeta -i\gamma)t(zq^{-2})\)
              \qq (\zeta )\qq '(\zeta -i\gamma )
            + \non\\& 
            + \Delta _{\gamma}^{-1}\(f(\zeta -i\gamma)t'(zq^{-2})\)
             \qq (\zeta -i\gamma )\qq '(\zeta )-\non\\
             &-\Delta _{\gamma}^{-1}\(f(\zeta)t(z)\)
              \qq (\zeta -i\gamma )\qq '(\zeta )\non\\
             &-\Delta _{\gamma}^{-1}\(f(\zeta)t'(z)\)
             \qq (\zeta )\qq '(\zeta -i\gamma )
             \bigr)+\non\\
            & +f(\zeta)\qq (\zeta -i\gamma )\qq '(\zeta -i\gamma )
             +f(\zeta -i\gamma)\qq (\zeta )\qq '(\zeta )\bigr\}\non
\end{align}
Define
$$
s^-_k(\zeta)=\left\{\matrix -s_k(z)+u[f](\zeta),\qquad\qquad\qquad\  f=z^{k-g-1},
\quad k\ge 1;\\ \ \\
-s_0(z)+u[f](\zeta)+(-1)^{g+1}2, \quad f=\zeta z^{-g-1},\ k=0;\\ \ \\
-z^{g+1+k},\qquad\qquad\qquad\qquad\qquad -g\le k\le -1 \endmatrix
\right.
$$
and
$$
p_k(\zeta)=\left\{\matrix v[f](\zeta),\quad f=z^{k-g-1},\ k\ge 1;\\ \ \\
v[f](\zeta),\quad f=\zeta z^{-g-1},\ k=0;\\ \ \\
0,\quad -g\le k\le -1 \endmatrix
\right.
$$
These definitions imply that 
\begin{align}
&(s_k(z)+s^-_k(z))\ \qq (\zeta )\qq '(\zeta )
=\delta _{\gamma}(p_k(\zeta ))
\label{pk}
\end{align}

Similarly one introduces the functions $S_k(Z)$,
$S_k^-(\zeta)$, $P_k(\zeta)$ changing everywhere $z$ by $Z$,
$q$ by $Q$ and $i\gamma $-shift of $\zeta$ by $i\pi$-shift of
$\zeta$. One has
\begin{align}
&(S_k(z)+S^-_k(z))\ \qq (\zeta )\qq '(\zeta )
=\delta _{\pi}(P_k(\zeta ))
\label{Pk}
\end{align}
Our goal is to define a pairing $\langle \ l\ |\ L\ \rangle $
between two arbitrary polynomials $l(z)$ and $L(Z)$
such that $l(0)=0$, $L(0)=0$.
Notice that every such polynomial $l$ ($L$)
can be presented as linear combination of 
polynomials $s_k$ ($S_k$).

Consider the following two pictures: 
\vskip 1cm
\hskip -0.5cm  
\epsffile{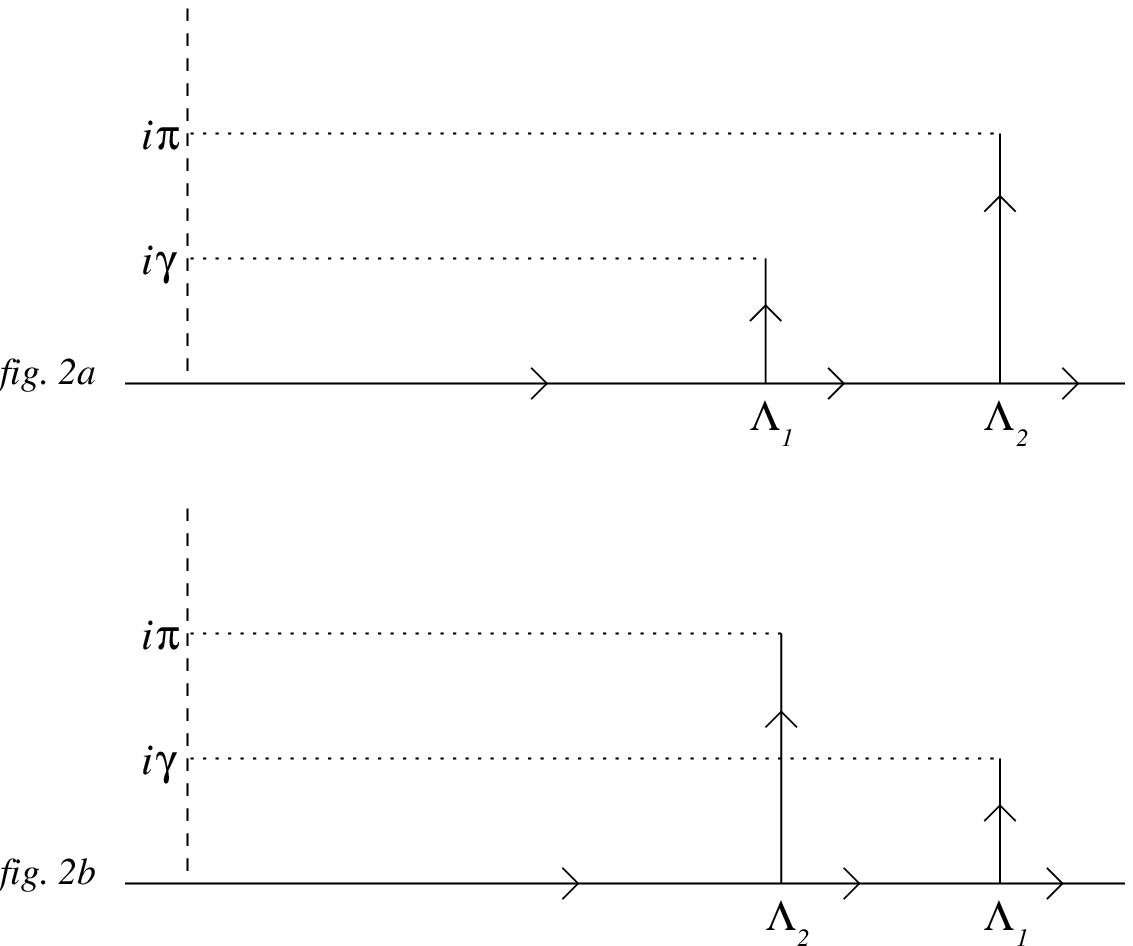}
\vskip 0.2cm
\noindent
We define
\begin{align}
&\langle s_k\ |\ S_l\rangle \equiv 
\int\limits _{-\infty}^{\Lambda _1}
\qq (\zeta )\qq '(\zeta )s_k(z)S_l(Z)d\zeta +
\non\\
&+\int\limits _ {\Lambda _1}^{\Lambda _2}
\qq (\zeta )\qq '(\zeta )s_k^-(\zeta )S_l(Z)d\zeta
+
\int\limits _ {\Lambda _2}^{\infty}
\qq (\zeta )\qq '(\zeta )s_k^-(\zeta )S_l^-(\zeta)d\zeta
-\non\\
&-\int\limits _{\Lambda _1}^{\Lambda _1+i\ga}
S_l(Z) p_k(\zeta )d\zeta 
-\int\limits _{\Lambda _2}^{\Lambda _2+i\pi}
s_k^-(\zeta) P_l(\zeta )d\zeta,
\label{reg}
\end{align}
see
({\it fig. 2a}).
The first integral in RHS converges at $-\infty $
because $l(0) =L(0)=0$.
The equations (\ref{pk},\ref{Pk})
guarantee that the regularization (\ref{reg})
does not depend on $\Lambda _1$, $\Lambda _2$ if they remain
ordered: $\Lambda _1 <\Lambda _2$.
Moreover, let us transform ({\it fig. 2a}) into ({\it fig. 2b}).
The alternative definition of regularized integral
referring to ({\it fig. 2b}) is 
\begin{align}
&\langle s_k\ |\ S_l\rangle \equiv 
\int\limits _{-\infty}^{\Lambda _2}
\qq (\zeta )\qq '(\zeta )s_k(z)S_l(Z)d\zeta +
\non\\
&+\int\limits _ {\Lambda _2}^{\Lambda _1}
\qq (\zeta )\qq '(\zeta )s_k(z)S_l^-(\zeta )d\zeta
+
\int\limits _ {\Lambda _1}^{\infty}
\qq (\zeta )\qq '(\zeta )s_k^-(\zeta )S_l^-(\zeta)d\zeta
-\non\\
&-\int\limits _{\Lambda _1}^{\Lambda _1+i\ga}
s_k(z) P_l(\zeta )d\zeta 
-\int\limits _{\Lambda _2}^{\Lambda _2+i\pi}
S_l^-(\zeta) p_k(\zeta )d\zeta
\label{reg'}
\end{align}
The equivalence of the regularizations (\ref{reg})
and (\ref{reg'}) is based on the following fact.
It is easy to realize that for any $l$ and $L$ there
exist a function $X _{kl}(\zeta )$ such that
\begin{align}
&(S_l(Z)-S_l^-(\zeta))p_k(\zeta )=\delta _{\pi}(X_{kl}(\zeta )),\non\\
&(s_k(z)-s_k^-(\zeta ))P_l(\zeta )=\delta _{\gamma}(X_{kl}(\zeta ))\non
\end{align}
The equivalence in question follows from the equality:
\begin{align}
&\int\limits _{\Lambda }^{\Lambda +i\gamma }
(S_l(Z)-S_l^-(\zeta))\ g(\zeta )d\zeta =
\(\int\limits _{\Lambda+i\pi }^{\Lambda +i\pi+i\gamma }-
\int\limits _{\Lambda }^{\Lambda +i\gamma}\ \)
\ X_{kl}(\zeta )d\zeta =\non\\
&
=
\(\int\limits _{\Lambda+i\gamma }^{\Lambda +i\pi+i\gamma }-
\int\limits _{\Lambda }^{\Lambda +i\pi}\ \)\ X_{kl}(\zeta )d\zeta =
\int\limits _{\Lambda }^{\Lambda +i\pi}
(s_k(z)-s_k^-(\zeta ))\ G(\zeta )d\zeta \non
\end{align}

\end{document}